\documentclass[aps,prl,twocolumn,showpacs,superscriptaddress,groupedaddress]{revtex4-1}
\usepackage{graphicx}  
\usepackage{dcolumn}   
\usepackage{bm}        
\usepackage[latin1]{inputenc}
\usepackage{amsmath}
\usepackage{amsfonts}
\usepackage{amssymb}
\usepackage{rotating}
\usepackage{subfigure}
\usepackage{paralist} 
\usepackage{verbatim}
\usepackage{color}
\usepackage{soul} 
\usepackage{appendix}

\usepackage{epsfig,hyperref}
\usepackage{array}
\usepackage{threeparttable}

\newcommand{\jcap}{JCAP}
\newcommand{\apjl}{ApJL}

\newcommand{\aj}{AJ}
\newcommand{\mnras}{MNRAS}

\definecolor{orange}{rgb}{1,0.3,0}

\newcommand{\mpch}{\>h^{-1}{\rm {Mpc}}}

\def\gcm3{\mathrm{g} / \mathrm{cm}^3}

\def\gtsima{$\; \buildrel > \over \sim \;$}
\def\ltsima{$\; \buildrel < \over \sim \;$}
\def\prosima{$\; \buildrel \propto \over \sim \;$}
\def\gsim{\lower.7ex\hbox{\gtsima}}
\def\lsim{\lower.7ex\hbox{\ltsima}}
\def\simgt{\lower.7ex\hbox{\gtsima}}
\def\simlt{\lower.7ex\hbox{\ltsima}}
\def\simpr{\lower.7ex\hbox{\prosima}}

\begin{document}
\title{Measurement of a Cosmographic Distance Ratio with Galaxy and CMB Lensing}

\author{Hironao Miyatake}
\affiliation{Jet Propulsion Laboratory, California Institute of Technology,
Pasadena, CA 91109, USA}
\affiliation{Kavli Institute for the Physics and Mathematics of the Universe (Kavli IPMU, WPI), The University of Tokyo, Chiba 277-8582, Japan}
\author{Mathew S. Madhavacheril}
\affiliation{Department of Astrophysical Sciences, 
  Princeton University, Princeton, NJ 08544 USA}
\affiliation{Physics and Astronomy Department, Stony Brook University, 
Stony Brook, NY 11794, USA}
\author{Neelima Sehgal}
\affiliation{Physics and Astronomy Department, Stony Brook University, 
Stony Brook, NY 11794, USA}
\author{An\v{z}e Slosar}
\affiliation{Physics Department, Brookhaven National Laboratory, 
Brookhaven, NY 11973, USA}
\author{David N. Spergel}
\affiliation{Department of Astrophysical Sciences, 
Princeton University, Princeton, NJ 08544 USA}
\author{Blake Sherwin}
\affiliation{Berkeley Center for Cosmological Physics, LBL and
Department of Physics, University of California, Berkeley, CA, 94720 USA}
\author{Alexander van Engelen}
\affiliation{Canadian Institute for Theoretical Astrophysics, University of
Toronto, Toronto, ON, Canada M5S 3H8}
\date{\today}

\begin{abstract}
We measure the gravitational lensing shear signal around dark matter halos hosting CMASS galaxies using light sources at $z\sim 1$ (background galaxies) and at the surface of last scattering at $z\sim 1100$ (the cosmic microwave background). The galaxy shear measurement uses data from the CFHTLenS survey, and the microwave background shear measurement uses data from the {\it Planck} satellite. The ratio of shears from these cross-correlations provides a purely geometric distance measurement across the longest possible cosmological lever arm. This is because the matter distribution around the halos, including uncertainties in galaxy bias and systematic errors such as miscentering, cancels in the ratio for halos in thin redshift slices.
We measure this distance ratio in three different redshift slices of the CMASS sample, and combine them to obtain a $17\%$ measurement of the distance ratio, $r=0.390^{+0.070}_{-0.062}$ at an effective redshift of $z=0.53$. This is consistent with the predicted ratio from the {\it Planck} best-fit $\Lambda$CDM cosmology of $r=0.419$.
\end{abstract}

\maketitle

\section{Introduction}
Cross-correlating optical weak lensing and cosmic microwave background (CMB) lensing is emerging as a powerful tool for measuring cosmological parameters and quantifying systematic uncertainties.  In particular, cross-correlations between optical and CMB lensing are sensitive to structure growth, and thus dark energy properties and modifications to General Relativity on large scales \cite{deref1,deref2,deref3,VallSPTDES}.  These cross-correlations can also isolate systematic effects such as, for example, multiplicative and photo-$z$ biases in optical weak lensing measurements \cite{Baxter16,Liu16,Schaan16}.  Recently cross-correlations using CMB lensing data from ACT, SPT, and {\it Planck} and optical lensing data from the CFHTLenS and DES surveys have been presented with detections of modest significance \cite{HandACT,MM15,LiuHill,OmuriHolder,Tomasso15,Pullen15,Kirk16,Baxter16}.  However, the precision of these measurements is expected to increase rapidly with newer data from, e.g., ACTPol, SPTpol, CMB-S4, HSC, DES, KiDS, and LSST.

In this work, we present the first measurement of a particularly useful cross-correlation between optical and CMB lensing:~the cosmographic distance ratio.  This measurement is obtained by measuring the gravitational lensing shear around a particular set of dark matter halos, first using background galaxies as the lensed source plane and then using the CMB as the lensed source plane.  Taking the ratio of these shear measurements results in a purely geometric distance measurement that is insensitive to the details of the mass distribution around the lensing halos, their galaxy bias, or potential miscentering \cite{Jain03,Taylor07,Kitching08,Collett12,Linder16}.  The ratio is given by
\begin{equation}
r = \frac{\gamma_t^o}{\gamma_t^c}\sim\frac{d_A(z^c)d_A(z^L,z^g)}{d_A(z^g)d_A(z^L,z^c)}
\end{equation}
where ${\gamma_t^o}$ and ${\gamma_t^c}$ are the optical and CMB tangential shear, $z^c$, $z^g$, and $z^L$ are the redshifts to the CMB, the background galaxy source plane, and the lensing structure respectively, $d_A(z)$ is the angular diameter distance at redshift $z$ and $d_A(z_L,z_g)$ is the angular diameter distance between redshifts $z_L$ and $z_g$ \cite{HuHolzVale,DasSpergel09}. This ratio decreases with redshift and the equation-of-state parameter $w$ and increases with the curvature $\Omega_k$ (see Fig.~\ref{fig:ratio_z}). This ratio has been measured previously when both source planes have been background galaxies with $z < 2.5$ \cite{Kitching07,Medez11,Taylor12,Kitching15,CollettData}. However, the advantage of using the CMB as the second source plane is that it provides the longest lever arm for distance ratios, which can result in an order of magnitude higher sensitivity to dark energy parameters \cite{HuHolzVale,DasSpergel09}.  
In this Letter, we present the first measurement of such a ratio using data from {\it Planck}, CFHTLenS, and the BOSS CMASS galaxy sample. 

\section{Data \& Method}
\subsection{The Lenses: BOSS CMASS Galaxies}

For the foreground lens sample, we use the CMASS selection of galaxies from the DR11 release of the BOSS spectroscopic survey. These mostly red galaxies constitute an approximately volume-limited selection of luminous galaxies from SDSS-III that span a redshift range of $0.4<z<0.7$. They are very often ($90$\%) at the center of their host halos \cite{White11} with masses of around $M_{200}=2\times 10^{13} M_{\odot}$, measured both from optical \cite{Miyatake15} and CMB lensing \cite{MM15}. As such, they are excellent tracers of massive halos that lens background sources.

In both the optical and CMB analyses, each CMASS lens galaxy is weighted as recommended in Section 2.4 of \cite{Anderson2014},

\begin{equation}
\label{eq:cmass_weight}
w_l = (w_{\rm noz}+w_{\rm cp}-1)w_{\rm see}w_{\rm star}
\end{equation}
to account for redshift failures ($w_{\rm noz}$), fiber collisions ($w_{\rm cp}$), effects of seeing ($w_{\rm see}$) and reduction in the ability of galaxy detection due to bright stars ($w_{\rm star}$). Since the same weights are used in both the CMB and optical analysis, the precise weighting scheme does not matter for the cosmological analysis. However, we use the recommended weighting scheme since we show halo profiles in Fig.~\ref{fig:tangential shear}. To reduce systematics associated with the width in redshift of the sample, we divide the sample into three redshift slices (see Table~\ref{tab:submsaples}) and perform the analysis separately in each redshift slice, combining the results only when calculating the final distance ratio at an effective redshift (see Results Section). For completeness, we also perform the analysis on the full sample in one wide redshift bin, but do not discuss cosmological inference from this.

\begin{table}[t] 
\begin{threeparttable}
\caption{Number of CMASS Galaxies Used}
\centering
\begin{tabular}{c c c c} 
Redshift & Galaxy Density & Optical & CMB \\ [0.5ex]
Range & (per arcmin$^2$) & Analysis & Analysis \\ [0.5ex]
\hline 
$0.43<z<0.51$ & 0.007 & 2,895 & 211,441 \\
$0.51<z<0.57$ & 0.007 & 2,896 & 213,497 \\
$0.57<z<0.7$ & 0.008 & 3,108 & 229,341 \\
[0.5ex]
\hline
$0.43<z<0.7$ & 0.021 & 8,899 & 654,279 \\
\hline
\end{tabular}
\label{tab:submsaples}
\end{threeparttable}
\end{table}

\subsection{Source Plane 1: CFHTLenS Galaxies}
We use the public CFHTLenS catalog~\cite{Heymans:2012,Erben:2013} for
calculating the optical tangential shear. The total area of the
CFHTLenS survey is 154 deg$^2$ in four distinct fields. The
overlapping area with the SDSS DR11 data is 105 deg$^2$ which contains
8,899 CMASS galaxies.

The catalog has galaxy shapes, which were measured by a Bayesian model-fitting method called {\it lens}fit \cite{Miller:2013}, and photometric-redshifts (photo-$z$s) which were estimated with the BPZ code \cite{Benitez:2000,Coe:2006} by using point-spread-function (PSF) matched photometry \cite{Hildebrandt:2012}. The effective number density of CFHTLenS source galaxies is 14~arcmin$^{-2}$.

The tangential shear in the $i$-th radial bin is measured by stacking galaxy shapes of lens-source pairs ($ls$);
\begin{equation}
  \label{eq:optical_lensing_measurement}
  \langle \gamma_t^o (R_i)\rangle = \frac{\sum_{ls \in R_i}
      w_{ls} e_t^{ls}}{\sum_{ls \in R_i} w_{ls}}.
\end{equation}
where $R_i$ is the distance from the center of the galaxy in the plane of the galaxy perpendicular to the line-of-sight, $e_t^{ls}$ is the tangential component of galaxy shapes, $w_{ls}$ is a weight which is the product of the CMASS galaxy weight $w_l$ given by Eq.~\eqref{eq:cmass_weight} and the inverse-variance weight for galaxy shapes $w_s$ provided by the CFHTLenS catalog that is estimated from the intrinsic galaxy shape and measurement error due to photon noise. Here the source galaxies are selected so that the best-fit photo-$z$ is greater than the lens redshift. The probability of such galaxies being foreground will be taken into account when calculating theoretical predictions, which is described later.

\begin{figure}[t]
\includegraphics[width=0.95\columnwidth]{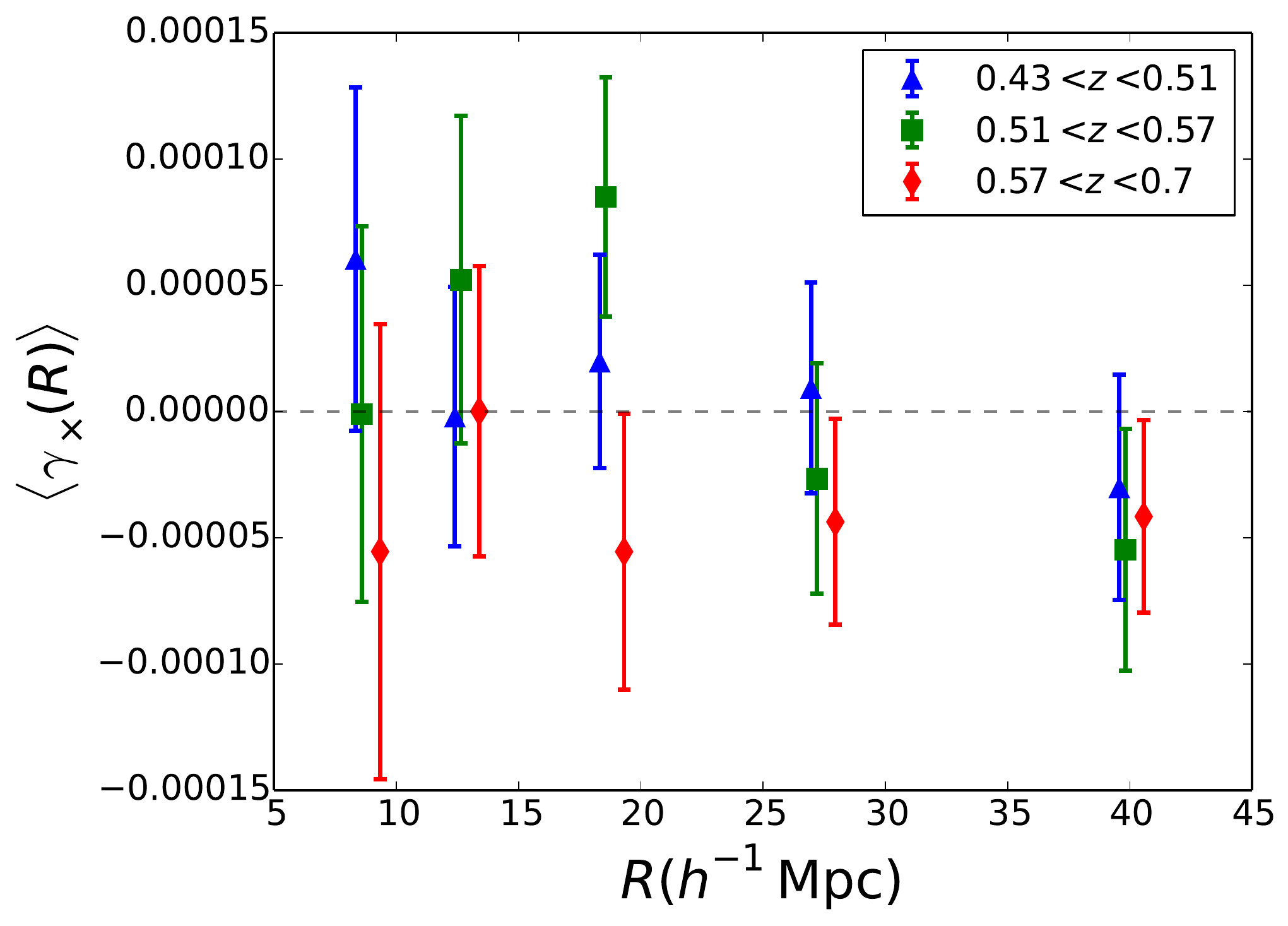}
\caption{Null test of optical lensing signal. The $R\sim 40\mpch$ bins are consistently smaller than zero for all the redshift slices, and thus we do not use them.  The $p$-value, or $P(\chi^2>\chi^2_{\rm obs})$ where $P$ is the $\chi^2$ distribution with degree of freedom of the 12 $R\simlt 30\mpch$ bins over the redshift slices, is 0.82, which is within a 95\%CL region. We use these 12 data points for the distance ratio analysis.}
\label{fig:optical_lensing_null_test}
\end{figure}

The covariance matrix of the tangential shear is estimated by measuring the tangential shear around 150 realistic mocks of the CMASS sample generated from $N$-body simulations \cite{manera13, manera15}. Using these CMASS mocks, we naturally include sample variance. Note that we could have canceled this sample variance exactly, by using exactly the same subset of galaxies to measure lensing of the CMB. However, given the large noise in the {\it Planck} convergence map, our overall statistical uncertainty would have increased.

If the PSF correction is imperfect, it can contaminate the tangential shear. To estimate this effect, we calculate the  tangential shear around random points. We use 50 realizations of random points to reduce statistical uncertainties \cite{Mandelbaum:2005}. The random signal is non-zero for $R \simgt 20\mpch$. We then make a PSF correction by subtracting this random signal from the lensing signal. If the correction works, the 45-degree-rotated shear should be consistent with zero. Figure~\ref{fig:optical_lensing_null_test} shows the 45-degree-rotated shear after the correction for each radial bin in each redshift slice. At the $R\sim 40\mpch$ radial bins, the PSF correction for tangential shear signal is as large as the statistical uncertainties of uncorrected signals in the first and second redshift slices. Thus we use the signal at $R\simlt 30\mpch$ for the distance ratio analysis.

\newcommand{\hp}{\texttt{HEALPIX }}

\subsection{Source Plane 2: Planck CMB Map}   
To extract a corresponding shear profile of CMASS halos using the CMB as the background light source, we prepare a \hp map \cite{healpix} of the CMASS galaxy overdensity (with $\texttt{nside}=1024$) for each redshift slice and cross-correlate it with the {\it Planck} reconstructed lensing convergence $\kappa$ map \cite{PlanckLens15}. Thus we obtain an estimate of $C_l^{\kappa\delta_g}$ in Fourier-space, which we then convert to a real-space shear estimate, $\langle \gamma_t^c(R)\rangle$, as discussed below.

\newcommand{\bx}{\mathbf{x}}

To create the galaxy overdensity map of CMASS galaxies, for each \hp pixel $\bx$, we assign a number given by 

\begin{equation}
\delta_g(\bx) = \frac{\sum_{i\in \bx} w_i}{\frac{1}{N}\sum_{i} w_i}-1
\end{equation}
where $\sum_{i\in \bx} w_i$ sums over the weights of each CMASS galaxy $i$ that falls in that pixel $\bx$, and where $\frac{1}{N}\sum_{i} w_i$ sums over the weights of all CMASS galaxies in all unmasked pixels and then divides by the total number of unmasked pixels $N$. Here the weight $w_i=w_lw_s(z)$, where $w_l$ is the BOSS systematic weight given in Eq.~\eqref{eq:cmass_weight} and $w_s(z)$ is an effective CFHTLenS weight. We include the CFHTLenS weights here, which have been interpolated as a function of lens redshift, because in the optical analysis they change the median redshift of the lens galaxies within a redshift slice. 

The mask used in this analysis is a combination of a mask derived from the completeness of the BOSS galaxies, where we exclude regions where the completeness is below 70\%, and the convergence mask provided with the {\it Planck} 2015 lensing data release. For the {\it Planck} convergence mask, we note that it masks out galaxy clusters identified through the thermal Sunyaev-Zeldovich effect.

We obtain a $C_l$ estimate of the cross-correlation by summing over spherical harmonic transform coefficients of the galaxy overdensity and CMB $\kappa$ maps, with the appropriate correction for fractional sky coverage ($f_{\mathrm{sky}}=0.206$ for 8,501 deg$^2$),

\begin{figure}[t]
\includegraphics[width=0.95\columnwidth]{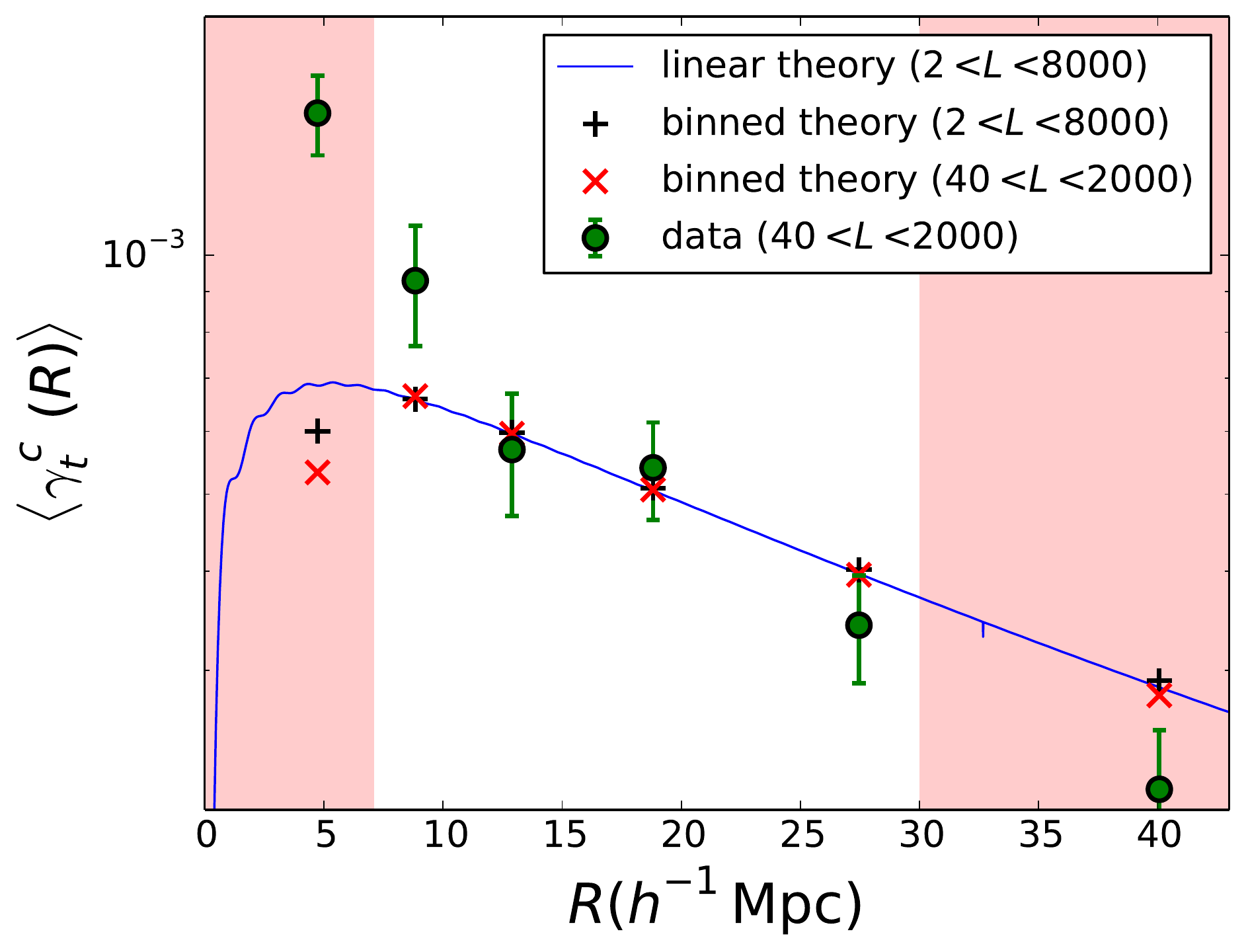}
\caption{Theory expectation of CMB tangential shear using an input $C_l^{\kappa\delta_g}$ curve generated with a linear matter power spectrum with a linear galaxy bias of 2. We do not use radial bins that have a mismatch between black crosses and red x's (shaded regions) as that would make the optical and CMB analyses inconsistent. The green points show the shear from the data, and where those points deviate from the theory at small scales is where there is sensitivity to the one-halo term from the CMASS galaxy halos themselves.}
\label{fig:2000vs8000}
\end{figure}

\begin{equation}
\hat{C}_l^{\kappa\delta_g} = \frac{1}{(2l+1)f_{\mathrm{sky}}^{\kappa\delta}}\sum_{m=-l}^{l} \delta_{lm}\kappa_{lm}.
\end{equation}
We then convert the cross-correlation estimate in Fourier-space to the real-space tangential shear of the CMB associated with CMASS galaxies, $\langle \gamma_t^c(R)\rangle$, via a Hankel transform (e.g Eq.~2 in \cite{HankelRef}),

\begin{equation}
\langle \gamma_t^c(R)\rangle = \frac{1}{2\pi}\int \ell d\ell J_2(\ell R/\chi) C_{\ell}^{\kappa\delta_g}.
\label{eq:Hankel}
\end{equation}
Note that this is exact only in the flat-sky limit, however we do not probe radial scales large enough that we should be sensitive to the effects of a curved sky.

To generate an expected theory curve we compute the shear transform in Eq.~\eqref{eq:Hankel} using an input $C_l^{\kappa\delta_g}$ curve generated with a linear matter power spectrum from \texttt{CAMB Sources} \cite{CAMBSources,halofit1,halofit2} with a linear galaxy bias of 2.  This is shown in Figure~\ref{fig:2000vs8000} both as the unbinned blue curve and as the black crosses binned identically to the data.  We also show here the result of restricting the $C_l^{\kappa\delta_g}$ to the range $40<L<2000$, which is the $L$ range of the {\it Planck} data $\kappa$-map used in this analysis. (Modes with $L<40$ can be affected by the treatment of the mask, and {\it Planck} does not report modes with $L>2048$). Including $2000<L<8000$ corresponds better to the resolution of the CFHTLenS survey, and in Figure~\ref{fig:2000vs8000} we show a significant difference at $R \sim 5\mpch$ between $L<2000$ and $L<8000$.  Thus we do not include this bin in our distance ratio analysis. We also exclude the radial bin at $R \sim 40\mpch$ because (a) it shows a small bias from the $L>40$ cut and (b) the same bin is excluded in the optical analysis as discussed previously.

We use 600 realizations of the CMASS mocks to make the covariance matrix and repeat the procedure above, cross-correlating a galaxy overdensity map generated from each mock with the {\it Planck} data $\kappa$-map, and then transforming that into a shear estimate. We note that there is no correlated structure between the {\it Planck} data $\kappa$-map and the CMASS mocks, so that the resulting covariance matrix does not include sample variance from this correlated structure.  However, this effect is expected to be negligible since the noise in the CMB $\kappa$-map is expected to dominate.  We check this by calculating Fisher-matrix theory errors with and without this $C_l^{\kappa\delta_g}$ term (see, e.g.,~Eq.~15 in \cite{Tomasso15}), and find agreement to within 1\% between the two.

\begin{figure}[t]
\includegraphics[width=0.95\columnwidth]{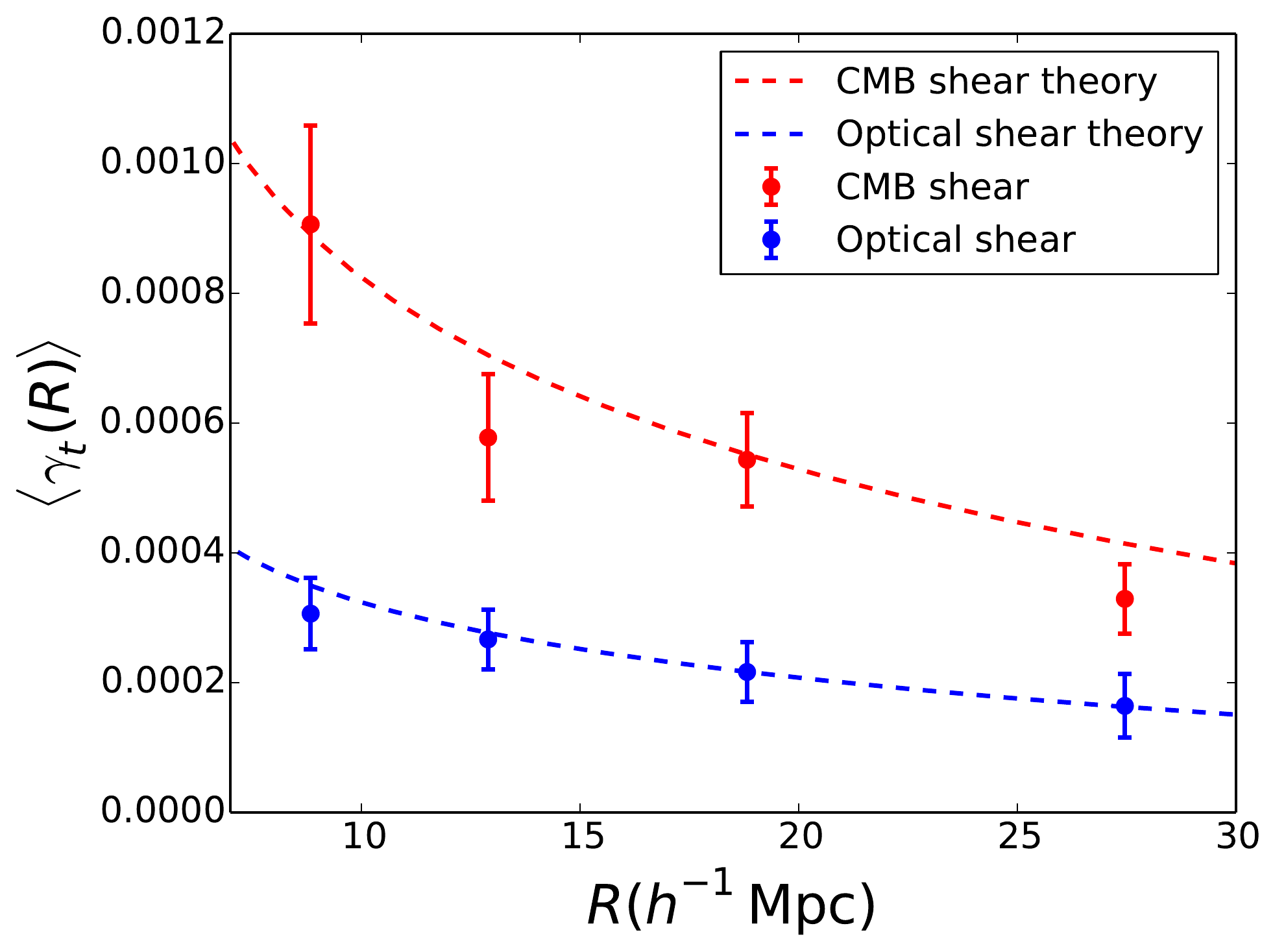}
\caption{CMB and optical shear around CMASS halos in the redshift range $0.43<z<0.7$. The dashed blue curve shows a theory fit to the optical data, which includes both the 1-halo and 2-halo terms.  This red curve is given by scaling up the blue curve to the CMB source redshift.}
\label{fig:tangential shear}
\end{figure}

\section{Results}
\label{sec:results}

Shear profiles, $\gamma_t(R)$, are related to the underlying projected mass density, $\Sigma(R)=\int d\chi \rho(R,\chi)$, through the relation 

\begin{equation}
\gamma_t(R)=\frac{\Delta\Sigma(R)}{\Sigma_{\rm cr}} = \frac{\bar{\Sigma}(<R)-\Sigma(R)}{\Sigma_{\rm cr}}
\end{equation}
where $\bar\Sigma(<R)$ is the average mass density within a circle of radius $R$, and $\Sigma_{\rm cr}$ is the critical surface mass density (defined below). We note that $\Delta\Sigma(R)$ depends only on the total matter distribution of the lens, and $\Sigma_{\rm cr}$ is a purely geometric quantity since it depends only on the distances to the lens and background sources. 
Since the criteria used to select the lensing galaxies
 is the same in the regions where the optical and CMB analyses are performed, we assume that the underlying $\Delta\Sigma(R)$ is identical in both cases. We also assume that the lensing by structures at higher redshift than the galaxy source is uncorrelated with the lower-redshift lensing.
This allows us to write the expected distance ratio as 
\begin{equation}
r(\{c_p\}) = \frac{\langle \gamma^o_t \rangle}{\langle \gamma^c_t\rangle}  = \frac{\Sigma_{\rm cr}^{\rm CMB}(\{c_p\})}{\Sigma_{\rm cr}^{\rm opt}(\{c_p\})}
\label{eq:ratio}
\end{equation}
where the dependence on the cosmological parameters, $\{c_p\}$, enters through the distance-redshift relations.  Here the numerator is the critical surface density for CMB lensing, which is calculated as

\begin{figure}[t]
\includegraphics[width=0.95\columnwidth]{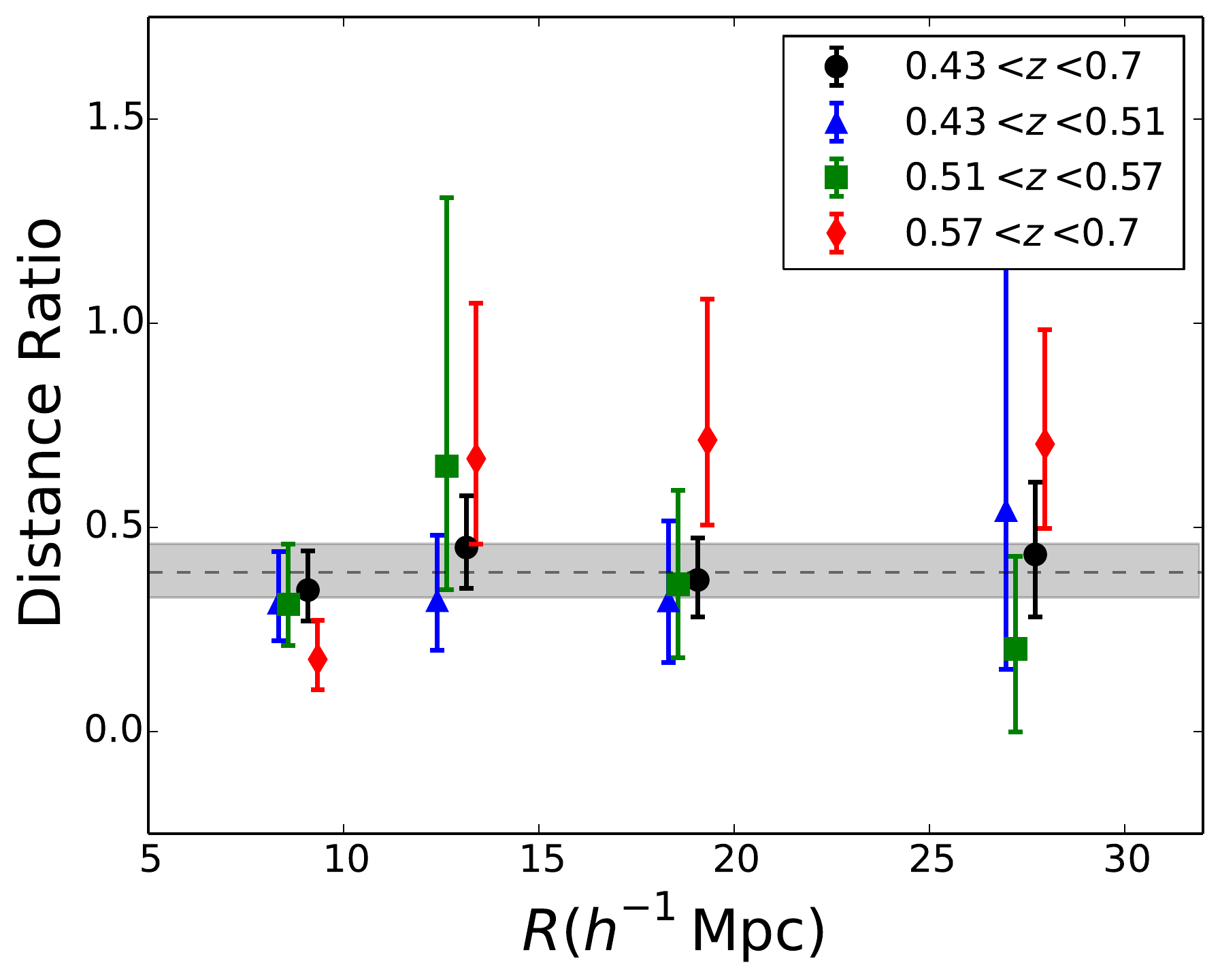}
\caption{Measured distance ratio for each radial bin and redshift slice of CMASS galaxies. Here the error bars are derived by Monte Carloing the covariance matrices for optical and CMB measurements, taking the ratio for each realization, and showing the 68\% CL region around the mean ratio. The dashed line and error band show $r=0.390^{+0.070}_{-0.062}$, the best-fit value coadding all the radial bins and simultaneously fitting to the three redshift slices.
}
\label{fig:12Ratios}
\end{figure}

\begin{equation}
\Sigma_{\rm cr}^{\rm CMB} = \left[ \frac{\sum_{ls} w_{l} P_{\rm stacked}(z_s|z_l) \Sigma_{\rm cr}^{-1}(z_l, z_{\rm CMB}; \{c_p\})}{\sum_{ls} w_{l}P_{\rm stacked}(z_s|z_l)} \right]^{-1}
\label{eq:sigma_cr_cmb}
\end{equation}
where $z_{\rm CMB}=1100$ 
, and the sum is over CMASS lenses. The critical surface density $\Sigma^{-1}_{\rm cr}$ is related to the angular diameter distances as,

\begin{equation}
\Sigma^{-1}_{\rm cr} = \frac{4\pi G}{c^2}\frac{d_A(z_l,z_s)d_A(z_l)(1+z_l)^2}{d_A(z_s)}.
\end{equation}
Here $d_A(z_s), d_A(z_l),$ and $d_A(z_l,z_s)$ are the angular diameter distances to the source, lens, and between the source and lens respectively. The $(1+z_l)^2$ factor comes from our use of comoving transverse separation $R$ in $\Delta\Sigma(R)$. We account for the weight dependence on the source galaxy redshift distribution in Eq.~\eqref{eq:optical_lensing_measurement}, using the photo-$z$ PDF stacked over optical source galaxies behind a given lens redshift;
\begin{equation}
P_{\rm stacked}(z|z_l) = \frac{\sum_s w_s P_s(z|z_l)}{\sum_s w_s}.
\end{equation}
The denominator in Eq.~\eqref{eq:ratio} is given by the equivalent expression for optical lensing. Note that the dilution effect due to foreground galaxies selected as source galaxies is effectively corrected for in the optical version of Eq.~\eqref{eq:sigma_cr_cmb}.

\begin{figure}[t]
\includegraphics[width=0.95\columnwidth]{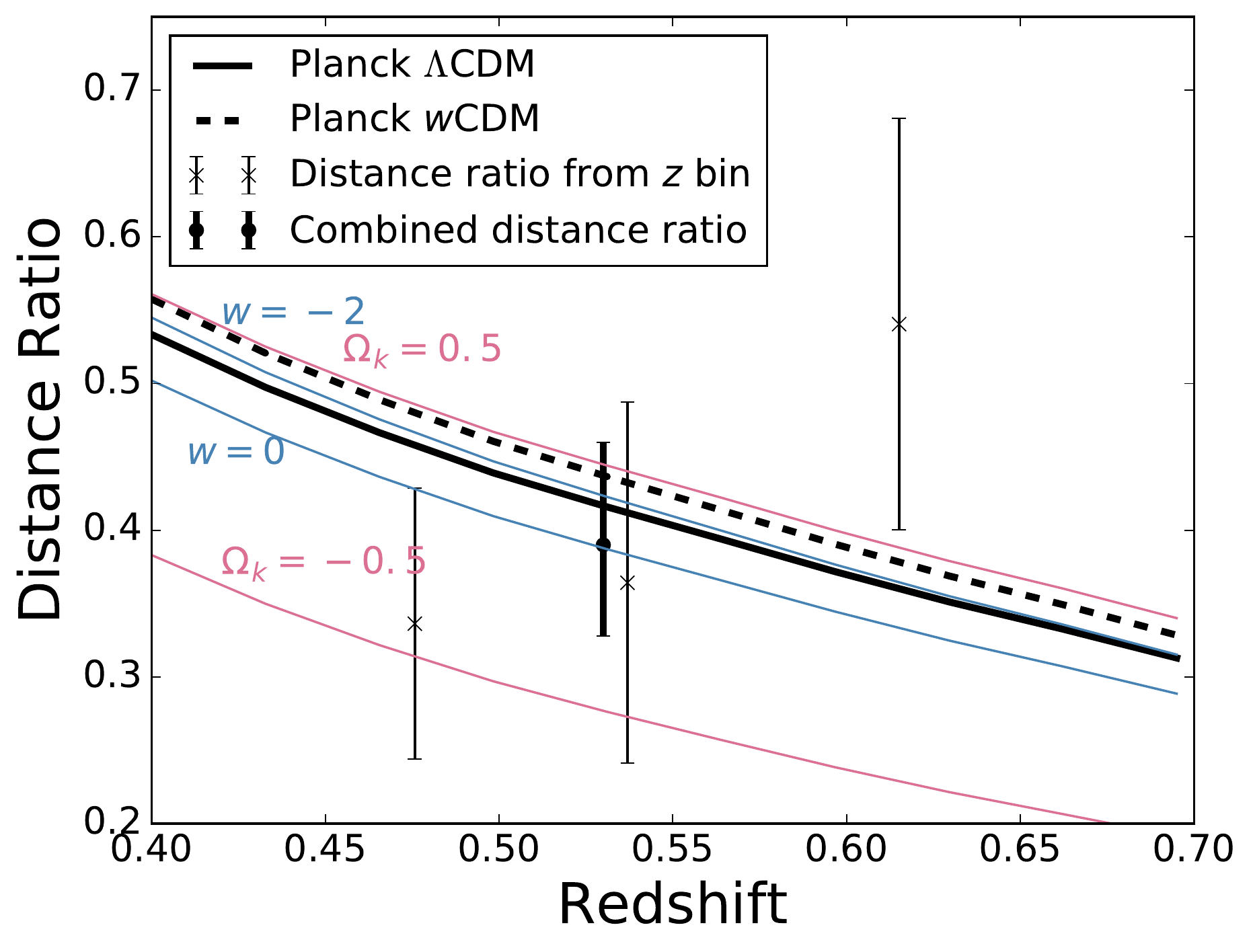}
\caption{Comparison of the measured distance ratio with that predicted from different cosmological models. 
The black solid and dashed curves show the ratio for the best-fit $\Lambda$CDM and $w$CDM models respectively from the {\it Planck} TT + lowP spectra \cite{PlanckParams15}. The thin solid curves show deviations from the best-fit {\it Planck} $\Lambda$CDM model as indicated.}
\label{fig:ratio_z}
\end{figure}

In Fig.~\ref{fig:tangential shear} and \ref{fig:12Ratios}, we show the measured tangential shear for the wide redshift slice and distance ratio for each radial bin and redshift slice of CMASS galaxies, respectively.
Fig.~\ref{fig:ratio_z} shows the coadded distance ratio for each redshift slice. We also include the distance ratio simultaneously fitted to the three redshift slices. In doing this, we assume the ratio linearly depends on redshift, i.e., $r(z|r_0, r^\prime)=r_0+r^\prime(z-z_{\rm p})$, 
and minimize the following quantity
\begin{equation}
\chi^2(r_0, r^\prime) = \sum_{\alpha}\sum_{ij}d_{i} {\rm Cov}^{-1}_{ij} d_{j},
\label{eq:chisq}
\end{equation}
where $d_i = \gamma^o (R_i) - r(z_{\alpha}|r_0, r^\prime)\gamma^c (R_i)$ for the $i$th radial bin, and where the covariance is given by
\begin{eqnarray}
{\rm Cov}_{ij} &=& {\rm Cov}(\gamma^o (R_i), \gamma^o (R_j)) \nonumber \\
&&- 2r {\rm Cov}(\gamma^o (R_i), \gamma^c (R_j)) \nonumber \\
&&+ r^2 {\rm Cov}(\gamma^c (R_i), \gamma^c (R_j)).
\label{eq:covmat}
\end{eqnarray}
We ignore the second term in Eq.~\eqref{eq:covmat} because the overlapping region for the two measurements is less than 2\% of the region used in our CMB analysis. The index $\alpha$ in Eq.~\eqref{eq:chisq} runs over the three redshift bins of the CMASS sample. Correlations between $z$-bins due to sample variance are not included because the contribution from clustering of CMASS galaxies was found to be subdominant to the contributions from CMB lensing reconstruction noise, Poisson noise of CMASS counts, and shape noise of CFHTLenS galaxies. The ``pivot'' redshift $z_{\rm p}$ is determined so that the errors on $r_0$ and $r'$ are uncorrelated. This yields $r=0.390^{+0.070}_{-0.062}$ at a pivot redshift of $z_{\rm p}=0.53$, a 17$\%$ measurement of distance ratio, and $r^\prime=1.2\pm0.97$. In Fig.~\ref{fig:ratio_z}, we also show the ratio predicted for different cosmological models as a function of lens redshift using Eq.~\eqref{eq:ratio}, assuming all the lenses are at a single redshift. Our measurement is consistent with the prediction of the {\it Planck} best-fit $\Lambda$CDM cosmology $r=0.419$ at $z_{\rm p}=0.53$ (within $1$-$\sigma$ statistical uncertainty). We obtain $\chi^2=3.15$ between the prediction and three redshift bins.

As potential systematic uncertainties of the optical shear analysis, we explore the effect of possible multiplicative shear bias $m$ and photo-$z$ bias $b_z$ on the optical measurement. To constrain these biases, we minimize 
Eq.~\eqref{eq:chisq} now replacing $(r_0, r^\prime) \rightarrow (m, b_z)$ and $d_i \rightarrow \gamma^o (R_i; m) - r(\{c_p\}, b_z)\gamma^c (R_i)$.
Since these biases affect the overall amplitude of the lensing signal, they are totally degenerate. Thus we investigate these biases separately. First, we parametrize multiplicative bias as $\gamma^o_{\rm obs} = (1+m) \gamma^o_{\rm true}$, and fit the distance ratio with cosmological parameters fixed to the {\it Planck} best-fit $\Lambda$CDM cosmology. The obtained constraint is $m=0.00^{+0.18}_{-0.16}$. Second, we parameterize the photo-$z$ bias as a shift of photo-$z$ PDF, i.e., $P(z)\rightarrow P(z-b_z)$. To avoid calculating the optical lensing signal with a new source galaxy selection every time $b_z$ is updated, we calculate the lensing signal without any source galaxy selection, which means all the dilution correction is put into $\Sigma_{\rm cr}^{\rm opt}$. 
With the fixed cosmology, we obtain $b_z=0.00^{+0.13}_{-0.12}$. These results indicate 
that there is no significant evidence of systematic uncertainties in our optical shear measurement. 

We also note that our analysis includes CMB lensing angular scales in the range $400<L<2000$, which region was excluded from the {\it Planck} lensing autospectrum analysis \cite{PlanckLens15}.  The reason for this exclusion was due to a failure of the curl null test around $L\sim700$.  While there may be a systematic affecting the autospectrum analysis, in general, one would expect many systematics to not be present in a cross-correlation analysis.

\section{Discussion}

In this work we have for the first time computed the distance ratio using optical and CMB weak lensing, yielding a 17\% measurement. We have used BOSS CMASS galaxies for the lensing galaxies, and CFHTLenS galaxy shapes and the {\it Planck} convergence map for optical and CMB background sources, respectively. The distance ratio extracts a purely geometrical factor by canceling out the matter distribution around halos, and thus we are free from systematic uncertainties arising from modeling galaxy bias and miscentering. We note that our separation of the lenses into thin redshift slices is critical for cancellation of the matter distribution in the ratio. Correlated structure along the line-of-sight will cause a slight difference from the calculation that assumes all lensing is done by the halo, but this is likely a very small effect. Our distance ratio is consistent with the predicted ratio from the {\it Planck} best-fit $\Lambda$CDM cosmology. Future optical lensing surveys (HSC, DES, KiDS, LSST, WFIRST and Euclid), combined with upcoming spectroscopic redshift surveys (PFS, DESI) and CMB surveys (AdvancedACT, SPT3G, the Simons Observatory, and CMB Stage-4) will allow for measurements of the distance ratio to within 1\% making it a competitive and complementary probe of curvature and cosmic acceleration.\\ 

\begin{acknowledgments}
We would like to thank Eric Linder for useful comments on the draft. HM would like to thank Shadab Alam and Rachel Mandelbaum for useful discussions. MM would like to thank Alexie Leauthaud and Paul Stankus for very useful conversations.  HM is supported in part by Japan Society for the Promotion of Science (JSPS) Research Fellowships for Young Scientists, by MEXT Grant-in-Aid for Scientific Research on Innovative Areas (No. 15H05893, 15H05892), and by the Jet Propulsion Laboratory, California Institute of Technology, under a contract with the National Aeronautics and Space Administration. NS acknowledges support from NSF grant number 1513618.

This work is based on observations obtained with MegaPrime/MegaCam, a joint project of CFHT and CEA/IRFU, at the Canada-France-Hawaii Telescope (CFHT) which is operated by the National Research Council (NRC) of Canada, the Institut National des Sciences de l'Univers of the Centre National de la Recherche Scientifique (CNRS) of France, and the University of Hawaii. This research used the facilities of the Canadian Astronomy Data Centre operated by the National Research Council of Canada with the support of the Canadian Space Agency. CFHTLenS data processing was made possible thanks to significant computing support from the NSERC Research Tools and Instruments grant program.
\end{acknowledgments}

\bibliographystyle{apsrev4-1}
\bibliography{main}

\end{document}